\documentclass[letterpaper,aps,prl,twocolumn,superscriptaddress,amsmath, amssymb,showpacs]{revtex4}

\newif\ifpdf
\ifx\pdfoutput\undefined
    \pdffalse           
\else
    \pdfoutput=1        
    \pdftrue
\fi
\ifpdf
    \usepackage[pdftex]{graphicx}
    \usepackage{epstopdf}
\else
    \usepackage{graphicx}
\fi
\usepackage{dcolumn}
\usepackage{bm}
\usepackage{setspace}
\usepackage{multirow}

\newcommand{\ppbar}{\ensuremath{p\bar{p}}}
\newcommand{\lnuqqbar}{\ensuremath{l{\bar{\nu}}q\bar{q}^{(\prime)}}}
\newcommand{\qqbarprime}{\ensuremath{q\bar{q}^\prime}}
\newcommand{\qqbar}{\ensuremath{q\bar{q}}}
\newcommand{\nunuqqbar}{\ensuremath{\nu\bar{\nu}q\bar{q}^{(\prime)}}}

\newcommand{\gevcc}{\ensuremath{\rm GeV/c}^{2}}

\newcommand{\invfb}{\ensuremath{\rm fb^{-1}}\,}

\newcommand{\measuredxsec}{\ensuremath{18.0\pm2.8\mathrm{(stat)}\pm2.4\mathrm{(syst)}\pm1.1\mathrm{(lumi)}}}
\newcommand{\measurednsig}{\ensuremath{1516\pm239\mathrm{(stat)}\pm144\mathrm{(syst)}}}

\def\simle{\mathrel{
    \rlap{\raise 0.511ex \hbox{$<$}}{\lower 0.511ex \hbox{$\sim$}}}}

\def\met{\mathbin{E\mkern - 11mu/_T}}

\def\metvec{\mathbin{\vec{E\mkern - 11mu/}_T}}
\def\mptvec{\mathbin{\vec{p\mkern - 8mu/}_T}}
\newcommand{\lw}[1]{\smash{\lower1.8ex\hbox{#1}}}

\begin{document}



\title{First Observation of Vector Boson Pairs in a Hadronic Final State at the Tevatron Collider}

\affiliation{Institute of Physics, Academia Sinica, Taipei, Taiwan 11529, Republic of China} 
\affiliation{Argonne National Laboratory, Argonne, Illinois 60439} 
\affiliation{University of Athens, 157 71 Athens, Greece} 
\affiliation{Institut de Fisica d'Altes Energies, Universitat Autonoma de Barcelona, E-08193, Bellaterra (Barcelona), Spain} 
\affiliation{Baylor University, Waco, Texas  76798} 
\affiliation{Istituto Nazionale di Fisica Nucleare Bologna, $^y$University of Bologna, I-40127 Bologna, Italy} 
\affiliation{Brandeis University, Waltham, Massachusetts 02254} 
\affiliation{University of California, Davis, Davis, California  95616} 
\affiliation{University of California, Los Angeles, Los Angeles, California  90024} 
\affiliation{University of California, San Diego, La Jolla, California  92093} 
\affiliation{University of California, Santa Barbara, Santa Barbara, California 93106} 
\affiliation{Instituto de Fisica de Cantabria, CSIC-University of Cantabria, 39005 Santander, Spain} 
\affiliation{Carnegie Mellon University, Pittsburgh, PA  15213} 
\affiliation{Enrico Fermi Institute, University of Chicago, Chicago, Illinois 60637}
\affiliation{Comenius University, 842 48 Bratislava, Slovakia; Institute of Experimental Physics, 040 01 Kosice, Slovakia} 
\affiliation{Joint Institute for Nuclear Research, RU-141980 Dubna, Russia} 
\affiliation{Duke University, Durham, North Carolina  27708} 
\affiliation{Fermi National Accelerator Laboratory, Batavia, Illinois 60510} 
\affiliation{University of Florida, Gainesville, Florida  32611} 
\affiliation{Laboratori Nazionali di Frascati, Istituto Nazionale di Fisica Nucleare, I-00044 Frascati, Italy} 
\affiliation{University of Geneva, CH-1211 Geneva 4, Switzerland} 
\affiliation{Glasgow University, Glasgow G12 8QQ, United Kingdom} 
\affiliation{Harvard University, Cambridge, Massachusetts 02138} 
\affiliation{Division of High Energy Physics, Department of Physics, University of Helsinki and Helsinki Institute of Physics, FIN-00014, Helsinki, Finland} 
\affiliation{University of Illinois, Urbana, Illinois 61801} 
\affiliation{The Johns Hopkins University, Baltimore, Maryland 21218} 
\affiliation{Institut f\"{u}r Experimentelle Kernphysik, Universit\"{a}t Karlsruhe, 76128 Karlsruhe, Germany} 
\affiliation{Center for High Energy Physics: Kyungpook National University, Daegu 702-701, Korea; Seoul National University, Seoul 151-742, Korea; Sungkyunkwan University, Suwon 440-746, Korea; Korea Institute of Science and Technology Information, Daejeon, 305-806, Korea; Chonnam National University, Gwangju,500-757, Korea; Chonbuk National University, Jeonju 561-756, Korea} 
\affiliation{Ernest Orlando Lawrence Berkeley National Laboratory, Berkeley, California 94720} 
\affiliation{University of Liverpool, Liverpool L69 7ZE, United Kingdom} 
\affiliation{University College London, London WC1E 6BT, United Kingdom} 
\affiliation{Centro de Investigaciones Energeticas Medioambientales y Tecnologicas, E-28040 Madrid, Spain} 
\affiliation{Massachusetts Institute of Technology, Cambridge, Massachusetts  02139} 
\affiliation{Institute of Particle Physics: McGill University, Montr\'{e}al, Qu\'{e}bec, Canada H3A~2T8; Simon Fraser University, Burnaby, British Columbia, Canada V5A~1S6; University of Toronto, Toronto, Ontario, Canada M5S~1A7; and TRIUMF, Vancouver, British Columbia, Canada V6T~2A3} 
\affiliation{University of Michigan, Ann Arbor, Michigan 48109} 
\affiliation{Michigan State University, East Lansing, Michigan  48824}
\affiliation{Institution for Theoretical and Experimental Physics, ITEP, Moscow 117259, Russia} 
\affiliation{University of New Mexico, Albuquerque, New Mexico 87131} 
\affiliation{Northwestern University, Evanston, Illinois  60208} 
\affiliation{The Ohio State University, Columbus, Ohio  43210} 
\affiliation{Okayama University, Okayama 700-8530, Japan} 
\affiliation{Osaka City University, Osaka 588, Japan} 
\affiliation{University of Oxford, Oxford OX1 3RH, United Kingdom} 
\affiliation{Istituto Nazionale di Fisica Nucleare, Sezione di Padova-Trento, $^z$University of Padova, I-35131 Padova, Italy} 
\affiliation{LPNHE, Universite Pierre et Marie Curie/IN2P3-CNRS, UMR7585, Paris, F-75252 France} 
\affiliation{University of Pennsylvania, Philadelphia, Pennsylvania 19104}
\affiliation{Istituto Nazionale di Fisica Nucleare Pisa, $^{aa}$University of Pisa, $^{bb}$University of Siena and $^{cc}$Scuola Normale Superiore, I-56127 Pisa, Italy} 
\affiliation{University of Pittsburgh, Pittsburgh, Pennsylvania 15260} 
\affiliation{Purdue University, West Lafayette, Indiana 47907} 
\affiliation{University of Rochester, Rochester, New York 14627} 
\affiliation{The Rockefeller University, New York, New York 10021} 
\affiliation{Istituto Nazionale di Fisica Nucleare, Sezione di Roma 1, $^{dd}$Sapienza Universit\`{a} di Roma, I-00185 Roma, Italy} 

\affiliation{Rutgers University, Piscataway, New Jersey 08855} 
\affiliation{Texas A\&M University, College Station, Texas 77843} 
\affiliation{Istituto Nazionale di Fisica Nucleare Trieste/Udine, I-34100 Trieste, $^{ee}$University of Trieste/Udine, I-33100 Udine, Italy} 
\affiliation{University of Tsukuba, Tsukuba, Ibaraki 305, Japan} 
\affiliation{Tufts University, Medford, Massachusetts 02155} 
\affiliation{Waseda University, Tokyo 169, Japan} 
\affiliation{Wayne State University, Detroit, Michigan  48201} 
\affiliation{University of Wisconsin, Madison, Wisconsin 53706} 
\affiliation{Yale University, New Haven, Connecticut 06520} 
\author{T.~Aaltonen}
\affiliation{Division of High Energy Physics, Department of Physics, University of Helsinki and Helsinki Institute of Physics, FIN-00014, Helsinki, Finland}
\author{J.~Adelman}
\affiliation{Enrico Fermi Institute, University of Chicago, Chicago, Illinois 60637}
\author{T.~Akimoto}
\affiliation{University of Tsukuba, Tsukuba, Ibaraki 305, Japan}
\author{B.~\'{A}lvarez~Gonz\'{a}lez$^t$}
\affiliation{Instituto de Fisica de Cantabria, CSIC-University of Cantabria, 39005 Santander, Spain}
\author{S.~Amerio$^z$}
\affiliation{Istituto Nazionale di Fisica Nucleare, Sezione di Padova-Trento, $^z$University of Padova, I-35131 Padova, Italy} 

\author{D.~Amidei}
\affiliation{University of Michigan, Ann Arbor, Michigan 48109}
\author{A.~Anastassov}
\affiliation{Northwestern University, Evanston, Illinois  60208}
\author{A.~Annovi}
\affiliation{Laboratori Nazionali di Frascati, Istituto Nazionale di Fisica Nucleare, I-00044 Frascati, Italy}
\author{J.~Antos}
\affiliation{Comenius University, 842 48 Bratislava, Slovakia; Institute of Experimental Physics, 040 01 Kosice, Slovakia}
\author{G.~Apollinari}
\affiliation{Fermi National Accelerator Laboratory, Batavia, Illinois 60510}
% ...

\collaboration{CDF Collaboration\footnote{With visitors from $^a$University of Massachusetts Amherst, Amherst, Massachusetts 01003,
$^b$Universiteit Antwerpen, B-2610 Antwerp, Belgium, 
$^c$University of Bristol, Bristol BS8 1TL, United Kingdom,
$^d$Chinese Academy of Sciences, Beijing 100864, China, 
$^e$Istituto Nazionale di Fisica Nucleare, Sezione di Cagliari, 09042 Monserrato (Cagliari), Italy,
$^f$University of California Irvine, Irvine, CA  92697, 
$^g$University of California Santa Cruz, Santa Cruz, CA  95064, 
$^h$Cornell University, Ithaca, NY  14853, 
$^i$University of Cyprus, Nicosia CY-1678, Cyprus, 
$^j$University College Dublin, Dublin 4, Ireland,
$^k$University of Edinburgh, Edinburgh EH9 3JZ, United Kingdom, 
$^l$University of Fukui, Fukui City, Fukui Prefecture, Japan 910-0017
$^m$Kinki University, Higashi-Osaka City, Japan 577-8502
$^n$Universidad Iberoamericana, Mexico D.F., Mexico,
$^o$University of Iowa, Iowa City, IA  52242,
$^p$Queen Mary, University of London, London, E1 4NS, England,
$^q$University of Manchester, Manchester M13 9PL, England, 
$^r$Nagasaki Institute of Applied Science, Nagasaki, Japan, 
$^s$University of Notre Dame, Notre Dame, IN 46556,
$^t$University de Oviedo, E-33007 Oviedo, Spain, 
$^u$Texas Tech University, Lubbock, TX  79609, 
$^v$IFIC(CSIC-Universitat de Valencia), 46071 Valencia, Spain,
$^w$University of Virginia, Charlottesville, VA  22904,
$^x$Bergische Universit\"at Wuppertal, 42097 Wuppertal, Germany,
$^{ff}$On leave from J.~Stefan Institute, Ljubljana, Slovenia, 
}}
\noaffiliation


\date{\today}

\begin{abstract}
We present the first observation in hadronic collisions of the electroweak production of vector
boson pairs ($VV$, $V$=$W,Z$) where  
one boson decays to a dijet final state . The data correspond to 3.5~$\invfb$ 
of integrated luminosity of $\ppbar$ collisions at $\sqrt{s}=1.96$\,TeV collected by the CDF~II 
detector at the Fermilab Tevatron. 
We observe $\measurednsig$ diboson candidate events and measure 
a cross section $\sigma(\ppbar$$\to$$VV$$+$$X$) of $\measuredxsec$\,pb, in agreement with the 
expectations of the standard model. 
\end{abstract}


\maketitle


The production of heavy gauge boson pairs ($WW,WZ$, or $ZZ$) in $\ppbar$ collisions has been observed in the
fully leptonic final states at the Fermilab Tevatron collider~\cite{cdf_VV,d0_VV}. 
Diboson production has not yet been conclusively observed in $p\bar{p}$ collisions in decay channels involving 
hadrons~\cite{oldCDFwwwz}; however, evidence for diboson decays into an $\lnuqqbar$ final state ($l=e,\mu,\tau$; 
$q=u,d,s,c,b$) has been recently presented by D\O~collaboration~\cite{lnujj_d0}.
 
Measurements of diboson production cross sections provide tests of the self-interactions of the gauge 
bosons. Deviations from the standard model (SM) prediction for the production rates could indicate new 
physics~\cite{hagiwara}. Furthermore, given that diboson production is topologically similar to associated 
Higgs boson production, $\ppbar$$\to$$VH$$+$$X$\,($V$=$W,Z$), the analysis techniques described in this 
Letter are important for Higgs boson searches. 

Here we present the first observation at a hadron collider of diboson production with one boson decaying into 
leptons and the other into hadrons. 
The analysis is performed on a sample of events with large transverse  
momentum imbalance ($\met$)~\cite{coordinate,missingEt} and two jets whose invariant mass can be reconstructed. 
This signature is sensitive to not only $\lnuqqbar$, but also to $\nunuqqbar$ decays because we do not explicitly 
require presence of identified charged leptons. The limited dijet  mass resolution results in a significant overlap of the 
$W$$\to$$\qqbarprime$ and the $Z$$\to$$\qqbar$ dijet mass peaks, and 
therefore the combination of the three diboson signals, $WW$, $WZ$, and $ZZ$, is considered.

%

We analyze a dataset of $\ppbar$ collisions corresponding to an integrated luminosity of 3.5\,\invfb
collected with the CDF~II detector at the center-of-mass energy of 1.96\,TeV. The CDF~II detector is 
described in detail elsewhere~\cite{CDF_detect_A}. Here the components that are relevant to this search 
are briefly discussed. Surrounding the beam pipe, there is a tracking system consisting of a silicon 
microstrip detector, a cylindrical drift chamber, and a solenoid that provides a 1.4\,T magnetic
field along the beam axis.
The central and plug calorimeters, which respectively cover the pseudorapidity regions of $|\mathrm\eta|$$<$1.1 
and 1.1$<$$|\mathrm\eta|$$<$3.6~\cite{coordinate}, surround the tracking system with a projective tower geometry. 
The calorimeters are composed of  inner electromagnetic and outer hadronic sections that consist of 
lead-scintillator and iron-scintillator, respectively.
In the central region, the calorimeter consists of 48 modules, segmented into towers of granularity
$\Delta \eta \times \Delta\phi \approx  0.1 \times 0.26$. The energy resolution of the central 
electromagnetic calorimeter for electrons is $\sigma(E_T)/E_T=13.5\%/\sqrt{E_T({\rm GeV})}\oplus 1.5\%$~\cite{cem},
while the energy resolution of the central hadron calorimeter for charged pions that do not 
interact in the electromagnetic section is $\sigma(E_T)/E_T=50\%/\sqrt{E_T({\rm GeV})}\oplus 3\%$~\cite{cwha},
where $E_T$ is the transverse energy~\cite{coordinate}. The plug electromagnetic (PEM) and plug hadron 
(PHA) calorimeters are identically segmented into 480 towers whose sizes are 
$\Delta\eta\times\Delta\phi\simeq 0.1\times 0.13$ for $|\eta|$$<$1.8 to 
$\Delta\eta\times\Delta\phi\simeq 0.6\times 0.26$ for the towers centered at $|\eta|$$=$3.3. The
corresponding PEM and PHA energy  
resolutions are $\sigma(E_T)/E_T=14.4\%/\sqrt{E_T({\rm GeV})}\oplus 0.7\%$ and 
$\sigma(E_T)/E_T=74\%/\sqrt{E_T({\rm GeV})}\oplus 4\%$, respectively~\cite{plug}. 
The wall hadron calorimeter (WHA) covers the gap in acceptance between the central and plug hadron 
calorimeters, corresponding to 0.7$<$$|\eta|$$<$1.3, with segmentation similar to that of the central calorimeter. The 
energy resolution of WHA is $\sigma(E_T)/E_T=75\%/\sqrt{E_T({\rm GeV})} \oplus 4\%$~\cite{cwha} 
for charged pions that do not interact in the electromagnetic section.
A system of Cherenkov counters~\cite{clc}, located around the beam pipe and inside the plug calorimeters, 
is used to measure the number of inelastic $\ppbar$ collisions per bunch crossing and thereby the luminosity.

The diboson signal ($WW,WZ$, and $ZZ$) is simulated using the {\sc pythia v6.2} Monte\,Carlo generator~\cite{pythia}. 
The most significant backgrounds to the diboson signal are $W$$(l\bar{\nu})$+jets, $Z$$(\nu\bar{\nu})$+jets, and 
QCD multijet production (in the following referred to as multijet background or MJB). Other less significant backgrounds 
include $Z$$(l\bar{l})$+jets, $t\bar{t}$, and single $t$-quark production. The $W$+jets backgrounds are simulated using 
the fixed-order matrix element generator {\sc alpgen v2.1}~\cite{alpgen} which is interfaced with {\sc pythia v6.3} to 
simulate parton showering and fragmentation, the underlying event, and additional $\ppbar$ interactions 
in the same bunch crossing. The $Z$+jets and  $t$-quark production processes are simulated with {\sc pythia v6.2}. The 
detector response in all Monte\,Carlo samples is modeled by a {\sc geant}-based CDF~II detector simulation~\cite{cdfsim}.
The MJB does not typically result in signatures of large intrinsic $\met$. However, when jet energy is not 
measured accurately an event may be reconstructed with large
$\met$ and pass the analysis selection criteria. Because of the large multijet production rate, this can still be a 
significant background in a $\met$+jets based analysis. The MJB is determined from the data. All 
other background predictions are normalized using next-to-leading-order (NLO) calculations for SM cross sections.

The selection of signal events proceeds as follows: a set of $\met$-based triggers select events with a variety of 
$\met$ and jet requirements. All these triggers have benefited significantly from the calorimeter trigger upgrade 
completed in 2007~\cite{l2cal}. The majority ($94\%$) of events satisfy the inclusive $\met$ trigger which requires 
$\met$$>$45\,GeV. Jets are reconstructed in the calorimeter using the {\sc jetclu} cone algorithm~\cite{jetclu} with a 
cone radius of 0.4 in ($\eta,\phi$) space. The energy measured by the calorimeter is corrected for effects that distort 
the true jet energy~\cite{JESNIM}. Such effects include the non-linear response of the calorimeter to particle energy, 
loss of energy in uninstrumented regions of the detector, spectator interactions, and energy radiated outside of the 
jet cone. We select events that have $\met$$>$60\,GeV~\cite{metcor} and exactly two jets with $E_{T}$$>$25\,GeV and 
$|\eta|$$<$2.0. This ensures a trigger efficiency of 96$\%$$\pm$2$\%$ on signal. We search for diboson production in 
the dijet mass range 40\,$<$$M_{jj}$$<$160\,$\gevcc$. The lower edge of the mass range is chosen to ensure 
that events are on the trigger efficiency plateau with respect to dijet mass. 
In addition to the requirements discussed above, the electromagnetic fraction of the 
total energy for each of the two jets is required to be less than 90$\%$ to ensure that electrons and 
photons are not counted as jets.

In order to suppress the MJB we use a $\met$ resolution model to distinguish true $\met$ originating
from undetected neutrinos from fake $\met$ due to jets that are not measured accurately. The $\met$ significance
is a dimensionless quantity based on the energy resolution of the jets, on soft unclustered particles, and on the
event topology; details of its definition can be found in Ref.~\cite{metmodel}. The $\met$ significance is typically 
low when $\met$ arises from mismeasurement. In addition to having a small significance, the $\metvec$ will often be 
aligned with a jet. We select events with $\met$ significance larger than 4 and azimuthal angle between $\metvec$ and 
nearest jet ($\Delta\phi_{\met}^{jet}$) greater than 0.4 radians.

Finally, we apply several requirements that suppress contamination due to cosmic-ray, beam-related, 
and other non-collision backgrounds. 
Events are required to have at least one reconstructed vertex formed by charged particle tracks.
The transverse energies of all calorimeter towers are calculated with respect to the $z$ position of the primary 
vertex with the largest $\sum p_T$ of associated tracks. The electromagnetic fraction of the total event energy 
has to be larger than 30$\%$ in order to reduce beam-related backgrounds. The arrival time of both leading jets 
as measured by the electromagnetic shower timing system~\cite{EMtiming} has to be consistent with the $p\bar{p}$ 
collision time. The remaining non-collision background has a smooth $M_{jj}$ distribution and accounts for less 
than 90 of the 44,910 selected events.

The shape and normalization of the MJB are determined from the data. A vector, $\mptvec$,
analogous to the calorimeter-based $\metvec$, is constructed from the vector sum of the transverse momenta  
of particles measured in the tracking system, and is largely uncorrelated to $\metvec$ for events where jets are 
not reconstructed accurately.
In the absence of $\met$ arising from mismeasurement in the calorimeter, the $\metvec$ and $\mptvec$ will be 
aligned in most events. The MJB is expected to be the dominant background component at larger values of 
$\Delta\phi$(${\metvec},{\mptvec}$). The dijet mass shape and normalization for the remaining MJB contribution in 
the sample is found by selecting events with $\Delta\phi$(${\metvec},{\mptvec}$)$>$1.0 and subtracting out the 
non-MJB backgrounds. The normalization is scaled up to account for the MJB contamination in the region 
$\Delta\phi$(${\metvec},{\mptvec}$)$<$1.0. The shape of the MJB is fit to an exponential in $M_{jj}$ 
to derive a dijet mass template. The MJB shapes of $M_{jj}$ and $\Delta\phi$(${\metvec},{\mptvec}$) distributions 
are verified with a large statistics MC sample.   

The signal extraction is performed using a minimization of the unbinned extended negative log likelihood with {\sc RooFit} 
program~\cite{roofit}. 
Three $M_{jj}$ template 
distributions are used in the fit: the first is $V$+jets and $t$-quark production (in the following referred to as
``electroweak'' (EWK) backgrounds) and is taken from Monte\,Carlo simulation; the second is the MJB template, where 
the slope and normalization are Gaussian constrained  to their previously measured values; the third template 
describes the signal. The signal shape is comprised of the $WW,~WZ$, and $ZZ$  distributions. This template is 
obtained from a Gaussian+polynomial fit to the signal Monte\,Carlo simulation where the mean and the width of
the Gaussian distribution are
linearly dependent on the jet energy scale (JES).


To assess the effect of systematic uncertainties on the measurement, we address separately two classes of sources:
those that affect the signal extraction procedure, and those that affect the signal acceptance in the cross section 
calculation. The signal extraction systematic uncertainties come from uncertainties in signal and background 
shapes. The shape uncertainties take into account the effect of jet energy resolution (JER), JES, 
MJB shape, and the shape of the EWK background. The jet energy scale, and the shape and the normalization 
of MJB, are treated as nuisance parameters in the fit and Gaussian constrained  to their independently 
measured values. These uncertainties are therefore accounted for in the statistical uncertainty of the 
extraction.

The shape uncertainty for the EWK background is determined by using $\gamma$+jets data~\cite{phoID} as an 
alternative background model in the $M_{jj}$ fit. All major non-MJB backgrounds include  a gauge boson accompanied 
by jets. There are similarities between the $\gamma$+jets and $V$+jets production; however, due largely to the mass 
difference between the $\gamma$ and the $W/Z$, the kinematics is not identical. To take this into account the 
$\gamma$+jets data are weighted by the ratio of the dijet mass distributions of the EWK background MC samples to 
$\gamma$+jets {\sc pythia} sample. 
We use these adjusted $\gamma$+jets data to determine a systematic uncertainty on the EWK $M_{jj}$ template.
Selection cuts applied to $\gamma$+jets events are not identical to those applied to the $\met$+jets sample. For 
example, the $Z$ decay into neutrinos will register as $\met$ in the detector, while the photon $E_T$ will be 
measured in the calorimeter. For this reason we cut on the vector sum of the photon $E_T$ and any $\met$ present 
in $\gamma$+jets events at 60\,GeV, treating this sum as analogous to $\met$ in $V$+jets events. A further 
consideration in the construction of the $\gamma$+jets template is the effect of $\gamma$$+$$V$ events, as these events 
will cause a peak in the $\gamma+$jets dijet mass distribution. We subtract this contribution using the $\gamma$$+$$V$ 
{\sc pythia} sample. Finally, we perform two signal extraction fits using the default EWK and $\gamma$+jets templates, 
respectively. The uncertainty due to the shape of the EWK background is then estimated as the difference in the results 
obtained from these two fits. The described method accounts for a combined effect of JES, JER, and modeling of jets in 
MC on the EWK $M_{jj}$ template. 


The uncertainty associated with the JES is the dominant source of systematic uncertainty on the
acceptance, and, therefore, the cross section. The JES affects several of the variables used in the
event selection. The effect of the JES systematic uncertainty is quantified by varying the jet energies in the
signal Monte\,Carlo simulation to account for the $\pm 1 \sigma$ variations of the JES. Other less significant sources of
systematic uncertainty that affect the measured cross section are jet energy resolution, initial and
final state radiation (ISR/FSR), and  parton distribution functions (PDF). A summary of all  
sources of systematic uncertainty is presented in Table~\ref{tab:sys}. 

The measured yields for signal and backgrounds are given in Table~\ref{tab:fit}. We extract $\measurednsig$ 
signal events which correspond to a cross section of $\measuredxsec$\,pb, in agreement with the SM prediction of 
16.8$\pm$0.5\,pb obtained using the {\sc mcfm~v5.4} program~\cite{campbell} with CTEQ6.1M PDFs~\cite{cteq6m}. Based 
on the MC simulation, the acceptances for the $WW$, $WZ$, and $ZZ$ production is 2.5$\%$, 2.6$\%$, and 2.9$\%$, 
respectively. In the calculation of the combined diboson cross section, we assume that each signal process 
contributes proportionally to its predicted SM cross section: 11.7~pb for $WW$, 3.6~pb for $WZ$, and 1.5~pb for $ZZ$.   
 


\begin{figure}[tp]\centering
  \includegraphics[width=0.5\textwidth]{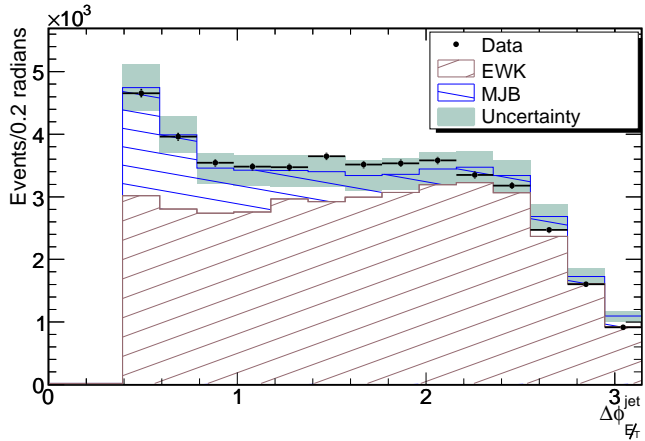}
  \caption{Data compared with the sum of the predicted EWK and MJB backgrounds for the
    $\Delta\phi^{jet}_{\met}$ variable. The band represents the total systematic uncertainty on the
    background. The measured signal is included here in the EWK contribution.} 
  \label{fig:dphiclosestwqcd}
\end{figure}

\begin{figure}[tp]
  \includegraphics[width=0.5\textwidth]{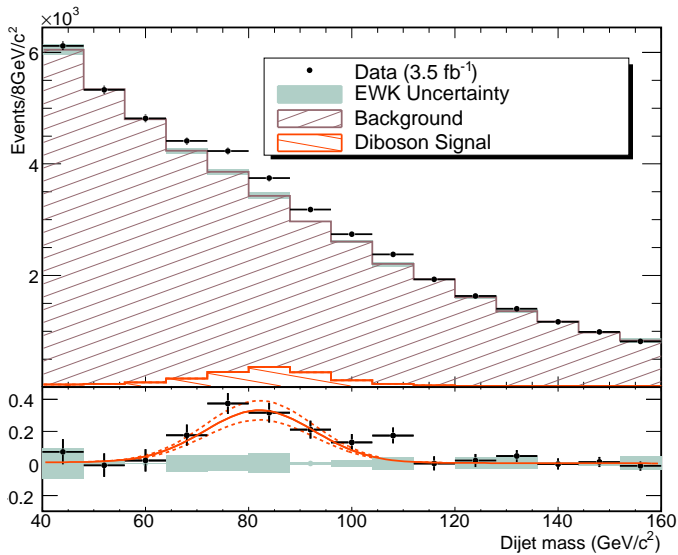}
  \caption{Top: Comparison between data and fitted background only. The measured signal is shown unstacked. 
The band represents the systematic uncertainty due to the shape of EWK background as described in text. 
Bottom: Comparison of the diboson signal (solid line) with the background-subtracted data (points). The 
dashed lines represent the $\pm 1\sigma$ statistical variations on the extracted signal. The gray band 
represents the systematic uncertainty due the EWK shape.}
  \label{fig:dijetmassfit}
\end{figure}

Figure~\ref{fig:dphiclosestwqcd} shows a comparison between the observed $\Delta\phi_{\met}^{jet}$ distribution
and the MJB and EWK (signal+background) components. This distribution provides a strong consistency check on our MJB model. 
Figure~\ref{fig:dijetmassfit} shows the fit result and a comparison between the expected signal and data after background 
subtraction. We bin the data as in Fig.~\ref{fig:dijetmassfit} and obtain a $\chi^2$ of 9.4 for 9 degrees of freedom 
corresponding to a p-value of $40\%$.

In summary, we use the $\met$+jets final state to measure the $WW$+$WZ$+$ZZ$ cross section  in $\ppbar$ collisions at 
$\sqrt{s}=1.96$ TeV to be $\measuredxsec$\,pb. This is consistent with the SM prediction of 16.8$\pm$0.5\,pb. 
To assess the strength of the observed signal, the effect of parameter variations due to all
relevant sources of uncertainty are studied 
by comparing the likelihood of background-only fit with the full fit result, and converting the difference into significance
numbers. We thus measure that the signal corresponds to a significance of at least 5.3 standard
deviations from the background-only hypothesis.


\squeezetable

\begin{table}[!ht]
\caption{The systematic uncertainties and their effect on the number of extracted signal events, the acceptance, 
and the cross section. All systematics are added in quadrature.}
\begin{center}
\begin{tabular}{cll}
        \hline \hline
& Systematic & \% uncert. \\ \hline
\multirow{2}{*}{\bf Extraction} &EWK shape   & 7.7 \\
&Resolution   & 5.6 \\ \hline
& {\bf Total extraction} & 9.5  \\ \hline \hline
\multirow{6}{*}{\bf Acceptance} & JES & 8.0    \\
& JER   & 0.7  \\
 & $\met$ resolution model & 1.0    \\
& Trigger inefficiency & 2.2 \\
& ISR/FSR & 2.5 \\
& PDF & 2.0 \\ \hline
& {\bf Total acceptance} & 9.0  \\ \hline 
& {\bf Luminosity} & 5.9 \\ \hline
& {\bf Total} & 14.4  \\ \hline \hline 
\end{tabular}
\label{tab:sys}
\end{center}
\end{table}

\begin{table}[hbt]
\begin{center}
\caption{Value of parameters in the model used to fit the $M_{jj}$ distribution. 1516$\pm$239(stat) signal
  events are extracted from the 44,910 data events which pass our selection cuts. The jet energy
  scale is also extracted from the fit and agrees well with the default value (1.0) as measured from calibrations~\cite{JESNIM}.} 
\begin{tabular}{ll} \hline \hline
Parameter & Fitted value\\ \hline
Jet energy scale, JES &  $0.985\pm0.019$ \\
Yield of EWK background events &  $36,140\pm1230$  \\
Yield of MJB events  &  $7249\pm1130$  \\
Yield of diboson candidates &  $1516\pm239$  \\ \hline \hline
\end{tabular}
\label{tab:fit}
\end{center}
\end{table}

\begin{acknowledgments}
%
We thank the Fermilab staff and the technical staffs of the
participating institutions for their vital contributions. This work
was supported by the U.S. Department of Energy and National Science
Foundation; the Italian Istituto Nazionale di Fisica Nucleare; the
Ministry of Education, Culture, Sports, Science and Technology of
Japan; the Natural Sciences and Engineering Research Council of
Canada; the National Science Council of the Republic of China; the
Swiss National Science Foundation; the A.P. Sloan Foundation; the
Bundesministerium f\"ur Bildung und Forschung, Germany; the Korean
Science and Engineering Foundation and the Korean Research Foundation;
the Science and Technology Facilities Council and the Royal Society,
UK; the Institut National de Physique Nucleaire et Physique des
Particules/CNRS; the Russian Foundation for Basic Research; the
Ministerio de Ciencia e Innovaci\'{o}n, and Programa
Consolider-Ingenio 2010, Spain; the Slovak R\&D Agency; and the
Academy of Finland. 
\end{acknowledgments}

\bibliography{metjjprl}

\begin{thebibliography}{43}
\bibitem[{coo()}]{coordinate}
\bibinfo{note}{ The detector is cylindrically
symmetric around the proton beam axis which is oriented in the positive $z$ direction.  The polar
angle, $\theta$, is measured from the origin of the coordinate system at the center of the detector
with respect to the $z$ axis. The pseudorapidity, transverse energy,
  and transverse momentum are defined as
  $\eta$=$-\ln\tan(\theta/2)$,
  $E_{T}$=$E\sin(\theta)$, and $p_{T}$=$p\sin(\theta)$, respectively.}
\bibitem[{metcor()}]{metcor}
\bibinfo{note}{In the offline event selection, we use $\met$ after correcting for 
jet energy losses in cracks between detector components and for non-linear calorimeter 
response~\cite{JESNIM}.}
\bibitem[{met()}]{metmodel}
\bibinfo{note}{
 The $\met$ significance is calculated as follows:
\begin{eqnarray}
\nonumber \met\mathrm{\,significance}=-\log_{10}(\widetilde{\mathcal{P}}), \\
\nonumber \widetilde{\mathcal{P}}=\prod\int_{-1}^{y_i}\mathcal{P}_{i}(x)dx,~~\mathrm{if}~~y_i<0,\\
\nonumber \mathrm{or}~~\widetilde{\mathcal{P}}=\prod\left(1-\int_{-1}^{y_i}\mathcal{P}_{i}(x)dx\right),~~\mathrm{if}~~y_i>0,\\
\nonumber y_i=\met/(E_{T}^{i}cos\Delta\phi_{i}),
\end{eqnarray}
\noindent where $\mathcal{P}_{i}(x)$ is the jet energy resolution function obtained from MC and validated with data, 
$E_{T}^{i}$ is the energy of the $i$-th jet, and $\Delta\phi_{i}$ is the azimuthal angle between the $i$-th jet and $\met$. }
\end{thebibliography}

\end{document}